\documentclass[10pt,aps,physrev,reprint,floatfix,superscriptaddress,longbibliography]{revtex4-2}
\pdfoutput=1
\usepackage{graphicx,latexsym}
\usepackage{dcolumn,amsfonts}
\usepackage{amssymb,amsmath,bm}

\usepackage[breaklinks=true,pdfencoding=auto]{hyperref}

\usepackage{natbib}
\usepackage{balance}

\hypersetup{
	colorlinks   = true, 
	urlcolor     = blue, 
	linkcolor    = blue, 
	citecolor    = red 
}

\usepackage{color}
\usepackage{ulem}

\begin{document}

	\title{Magneto-optical properties of a quantum dot array\\ interacting with a far-infrared photon mode of a cylindrical cavity}

\author{Vidar Gudmundsson}
	\email{vidar@hi.is}
	\affiliation{Science Institute, University of Iceland, Dunhaga 3, IS-107 Reykjavik, Iceland}
	\author{Vram Mughnetsyan}
	\email{vram@ysu.am}
	\affiliation{Department of Solid State Physics, Yerevan State University, Alex Manoogian 1, 0025 Yerevan, Armenia}
	\author{Hsi-Sheng Goan}
	\email{goan@phys.ntu.edu.tw}
    \affiliation{Department of Physics and Center for Theoretical Physics, National Taiwan University, Taipei 106319, Taiwan}
	\affiliation{Center for Quantum Science and Engineering, National Taiwan University, Taipei 106319, Taiwan}
	\affiliation{Physics Division, National Center for Theoretical Sciences, Taipei 106319, Taiwan}
	\author{Jeng-Da Chai}
	\email{jdchai@phys.ntu.edu.tw}
	\affiliation{Department of Physics and Center for Theoretical Physics, National Taiwan University, Taipei 106319, Taiwan}
	\affiliation{Center for Quantum Science and Engineering, National Taiwan University, Taipei 106319, Taiwan}
	\affiliation{Physics Division, National Center for Theoretical Sciences, Taipei 106319, Taiwan}
	\author{Nzar Rauf Abdullah}
	\email{nzar.r.abdullah@gmail.com}
	\affiliation{Physics Department, College of Science,
		University of Sulaimani, Kurdistan Region, Iraq}
	\author{Chi-Shung Tang}
	\email{cstang@nuu.edu.tw}
	\affiliation{Department of Mechanical Engineering, National United University, Miaoli 36003, Taiwan}
	\author{Valeriu Moldoveanu}
	\email{valim@infim.ro}
	\affiliation{National Institute of Materials Physics, PO Box MG-7, Bucharest-Magurele,
		Romania}
	\author{Andrei Manolescu}
	\email{manoles@ru.is}
	\affiliation{Department of Engineering, Reykjavik University, Menntavegur
		1, IS-102 Reykjavik, Iceland}

%

\begin{abstract}
We model the equilibrium properties of a two-dimensional electron gas in a square lateral superlattice
of quantum dots in a GaAs heterostructure subject to an external homogeneous perpendicular magnetic field
and a far-infrared circular cylindrical photon cavity with one quantized mode, the TE011 mode.
In a truncated linear basis constructed by a tensor product  of the single-electron states of the noninteracting
system and the eigenstates of the photon number operator, a local spin density approximation of density functional
theory is used to compute the electron-photon states of the two-dimensional electron gas in the cavity.
The common spatial symmetry of the vector fields for the external magnetic field and the cavity photon field
in the long wavelength approximation enhances higher order magnetic single- and multi-photon processes
for both the para- and the diamagnetic electron-photon interactions. The electron-photon coupling introduces
explicit photon replicas into the bandstructure and all subbands gain a photon content,
constant for each subband, that can deviate from an integer value as the coupling is increased or the photon energy
is varied. The subbands show a complex Rabi anticrossing behavior when the photon energy and the coupling
bring subbands into resonances. The complicated energy subband structure leads to photon density variations
in reciprocal space when resonances occur in the spectrum. The electron-photon coupling polarizes
the charge density and tends to reduce the Coulomb exchange effects as the coupling strength increases.

\end{abstract}

\maketitle
%
%

\section{Introduction}
Photon cavities have been proposed and used to tune or enhance the properties of electron and
material systems in the fields of chemistry \cite{https://doi.org/10.1002/anie.201107033-2,Davidsson2020,doi:10.1021/acs.jctc.8b00580,doi:10.1021/acsphotonics.7b01279,Mandal2023},
physics \cite{PhysRevB.99.035129,doi:10.1021/acs.jpclett.3c01294,doi:10.1073/pnas.1518224112} 
and material science \cite{Huebener2021}.
H{\"u}bener et al.\ have suggested engineering quantum materials with chiral optical cavities
to break the symmetry of the original system in order to obtain novel characteristics
\cite{Huebener2021}, and Wang et al.\ presented how cavity photon dynamics could be manipulated
by topologically curved space \cite{Wang2022}.

The extraordinary high polarizability and mobility of a two-dimensional electron gas (2DEG) in a GaAs
heterostructure make it an ideal experimental system for attaining nonperturbative coupling of electrons
with far-infrared (FIR) cavity photons \cite{Zhang1005:2016}. In a modulated 2DEG in a high-quality-factor
terahertz cavity in a magnetic field, the quasi-particles are Landau subband polaritons.

Nonrelativistic quantum electrodynamics, often for a single cavity-photon mode, together with
different approaches to the electron dynamics has been used to describe electronic systems
in photon cavities. For few electrons in a nanoscale system various toy-models have been
used with an emphasis on the electron-photon interaction
\cite{PhysRev.93.99,Jaynes63:89,10.1063/5.0076485,Yuan2017}.
In addition, models where both
the para- and the diamagnetic electron-photon interactions are included together with the 
Coulomb electron-electron interaction within a numerical exact diagonalization formalism
have been used to describe the properties of 
closed \cite{PhysRevE.86.046701,ANDP:ANDP201500298}
and open systems \cite{2016arXiv161003223J,GUDMUNDSSON20181672}. Larger electron systems
have commonly been modeled using some variant of Quantum Electrodynamical Density Functional 
Theory (QEDFT) \cite{doi:10.1073/pnas.1518224112,PhysRevA.98.043801,Rubio2021:2110464118,flick2021simple,PhysRevB.106.115308,PhysRevB.108.115306,10.1063/5.0123909}.

For a 2DEG in an external homogeneous magnetic field and a periodic
superlattice potential, the persistent equilibrium currents are rotational
\cite{Merkt96:1134,Grundler98:693}. In such a system, active magnetical
transitions, i.e.\ magnetic dipolar and higher order transitions, are of importance to influence
or control its properties via cavity photons.
One way to couple the cavity-photons to the predominantly rotational, or transverse, currents in the 2DEG is to
use a circular cylindrical cavity. In a far-infrared cavity, where the long wavelength approximation
is applicable as the wavelength of the photon field is much longer than the characteristic length
scale, the superlattice length $L$, the TE$_{011}$ cavity mode can play a
special role as its vector field has the same spatial symmetry as the vector field of the external
static homogeneous magnetic field. This choice of a cavity mode is thus selected not to
break the symmetry of the 2DEG, but rather to enhance the coupling of the matter-photon system
using its magnetically active processes stemming from both the para- and the diamagnetic
electron-photon interactions.

Utyushev et al.\ have recently discussed the generation of highly directional ``magnetic light''
from rare earth ions placed in, or near to, dielectric homogeneous spheres to enhance
magnetically active processes in the system \cite{utyushev2022generation}.

Our model calculations are based on a DFT approach for the electrons in a superlattice of quantum dots
and simultaneously bear a closeness to exact diagonalization, or configuration interactions (CI),
for photons in a cavity.
The calculations are performed in a basis constructed by single-electron states
of the noninteracting Hamiltonian tensor multiplied by the states of the photon number operator.
 Both the para- and the diamagnetic parts of the electron-photon
interactions are included in the long wavelength limit in order to include higher order virtual and 
real photon processes and vacuum effects consistently
\cite{Schafer2020,2017arXiv170603483G,Frisk-Kockum2019}.
The method using a linear space of a tensor product of electron and photon states in a DFT approach
mirrors what Malave et al.\ call QED-DFT-TP, quantum electrodynamics
\cite{10.1063/5.0123909}. Weight et al.\ investigated molecular exciton polaritons using a
similar approach and compared their results to those of a method where the electron-photon interaction is
not included in the self-consistency iterations, but added at their end \cite{doi:10.1021/acs.jpclett.3c01294}.
We have chosen the self-consistent QED-DFT-TP approach as we are dealing with Landau-subband polaritons
in the FIR regime.

The paper is organized as follows: In Sec.\ \ref{Model} we describe
the model. The results and discussion thereof are found in
Sec.\ \ref{Results}, with the conclusions drawn in Sec.\ \ref{Conclusions}.

\section{Model}
\label{Model}
We consider a 2DEG in a square lateral superlattice of quantum dots in a
GaAs heterostructure subject to a homogeneous external magnetic field.
The electrons have the effective mass $m^*=0.067m_e$,
the dielectric constant $\kappa = 12.4$, and the effective g-factor
$g^* = -0.44$.
The Hamiltonian of the 2DEG-cavity system in the photon cavity is
\begin{equation}
	H = H_\mathrm{e} + H_\mathrm{int} + H_\gamma,
\label{Hinitial}
\end{equation}
where
\begin{equation}
	H_\mathrm{e} = H_0 + H_\mathrm{Zee} + V_\mathrm{H} + V_\mathrm{per} + V_\mathrm{xc},
	\label{He}
\end{equation}
describes the 2D electrons in an array of quantum dots and
\begin{equation}
 	H_0 = \frac{1}{2m^*}\bm{\pi}^2, \quad\mbox{with}\quad
 	\bm{\pi} = \left(\bm{p}+\frac{e}{c}\bm{A} \right).
 	\label{H0}
\end{equation}
The vector potential ${\bm A} = (B/2)(-y,x)$ leads to the homogeneous
external magnetic field perpendicular to the plane of the 2DEG, $\bm{B}=B{\bm e}_z$.
The spin Zeeman term is $H_\mathrm{Zee} = \pm g^* \mu_\textrm{B}^* B/2$, and the direct Coulomb
interaction is
\begin{equation}
	V_\mathrm{H}(\bm{r}) = \frac{e^2}{\kappa}\int_{\mathbf{R}^2}d\bm{r}'\frac{\Delta n(\bm{r}')}
	{|\bm{r}-\bm{r}'|}
	\label{Vcoul}
\end{equation}
with $\Delta n(\bm{r}) = n_\mathrm{e}(\bm{r})-n_\mathrm{b}$, where $+en_\mathrm{b}$ is the
homogeneous positive background charge density reflecting the charge neutrality of the
total system. The electron charge density is $-en_\mathrm{e}(\bm{r})$, and
$\mu_\textrm{B}^*$ is the effective Bohr magneton.
The array of quantum dots is represented by the periodic potential
\begin{equation}
	V_\mathrm{per}(\bm{r}) = -V_0\left[\sin \left(\frac{g_1x}{2} \right)
	\sin\left(\frac{g_2y}{2}\right) \right]^2
	\label{Vper}
\end{equation}
with $V_0 = 16.0$ meV that defines the superlattice vectors
$\bm{R}=n\bm{l}_1+m\bm{l}_2$ with $n,m\in \bm{Z}$.
The unit vectors of the superlattice are $\bm{l}_1 = L\bm{e}_x$ and $\bm{l}_2 = L\bm{e}_y$,
and the inverse/reciprocal lattice is spanned by $\bm{G} = G_1\bm{g}_1 + G_2\bm{g}_2$ with
$G_1, G_2\in \mathbf{Z}$ and the unit vectors
\begin{equation}
	\bm{g}_1 = \frac{2\pi\bm{e}_x}{L}, \quad\mbox{and}\quad
	\bm{g}_2 = \frac{2\pi\bm{e}_y}{L}.
\end{equation}
The superlattice period is $L = 100$ nm. The derivation of the local spin density approximation (LSDA)
exchange and correlation potentials $V_\mathrm{xc}$ is documented in Appendix A of
Ref.\ \cite{PhysRevB.106.115308}. The interaction of the electrons with the vector potential,
${\bm A}_\gamma$, of the photon cavity in terms of the electron current, and charge densities is
\begin{align}
      H_\mathrm{int} = \frac{1}{c}\int_{\mathbf{R}^2} d\bm{r}\; &
      {\bm J}({\bm r})\cdot{\bm A}_\gamma (\bm{r}) \nonumber\\
      +& \frac{e^2}{2m^*c}\int_{\mathbf{R}^2} d\bm{r}\;
      n_\mathrm{e}(\bm{r})A^2_\gamma(\bm{r}).
\label{e-g}
\end{align}
In Appendix \ref{e-TE011} the electron-photon interaction (\ref{e-g}) for a single quantized TE$_{011}$ mode of
a cylindrical cavity is derived in the long wave approximation, i.e.\ when the spatial variation of the far-infrared
cavity field is only slight with respect to $L$. Formally, the interaction takes the form
\begin{align}
      H_\mathrm{int} &= g_\gamma \hbar\omega_c \left\{ lI_x + lI_y\right\} \left(a^\dagger_\gamma + a_\gamma\right)\nonumber\\
                     &+ g^2_\gamma \hbar\omega_c {\cal N}\left\{\left(a^\dagger_\gamma a_\gamma + \frac{1}{2}\right)
                                        +\frac{1}{2}\left(a^\dagger_\gamma a^\dagger_\gamma + a_\gamma a_\gamma\right)\right\}
\label{e-gIxIyN}
\end{align}
with the integrals, $I_x$, $I_y$, and ${\cal N}$ defined in Appendix \ref{e-TE011}. The dimensionless coupling
strength is
\begin{equation}
      g_\gamma = \left\{ \left( \frac{e{\cal A}_\gamma}{c} \right) \frac{l}{\hbar} \right\},
\end{equation}
while $\omega_c = eB/(m^*c)$ is the cyclotron frequency and $l = (\hbar c/(eB))^{1/2}$ is the magnetic length.
$a^\dagger_\gamma$ and $a_\gamma$ are the creation and annihilation operators for the photon mode with
fundamental energy $\hbar\omega_\gamma$ and the free Hamiltonian
\begin{equation}
      H_\gamma = \hbar\omega_\gamma a^\dagger_\gamma a_\gamma,
\end{equation}
where the zero point energy of the photon mode is neglected. Importantly, as is shown in Appendix \ref{e-TE011}
the vector potential of the cavity photon mode in the long wavelength approximation is
\begin{equation}
      {\bm A}_\gamma (\bm{r}) = \bm{e}_\phi {\cal A}_\gamma\left(a^\dagger_\gamma + a_\gamma  \right)
									   \left(\frac{r}{l} \right),
\end{equation}
where $\bm{e}_\phi$ is the unit angular vector in polar coordinates. This vector potential
happens to have the same spatial form as the vector potential ${\bm A}$ determining the
external homogeneous magnetic field $\bm{ B} = B\bm{e}_z$. This observation gives a natural scale
for the dimensionless coupling constant $g_\gamma$ as the strength of the spatial part of $\bm{A}_\gamma$
becomes equal to the magnitude of $\bm{A}$ when $g_\gamma = 1/2$.

We adapt a quantum electrodynamical density functional theory approach, QED-DFT-TP recently presented by Malave
\cite{10.1063/5.0123909} to our 2DEG-cavity system by calculating the energy spectrum and the eigenstates of
$H$ (\ref{Hinitial}) in a linear functional basis constructed by a tensor product (TP) of electron and photon states
\begin{equation}
         |\bm{\alpha\theta}\sigma n\rangle = |\bm{\alpha\theta}\sigma\rangle\otimes|n\rangle,
\label{TP}
\end{equation}
where the photon states are the eigenstates
of the photon number operator, and the electron states are the single electron states of Ferrari designed
for a periodic 2DEG in an external magnetic field at each point in the first Brillouin zone, i.e.\
$\bm{\theta} = (\theta_1,\theta_2)\in [-\pi,\pi]\times[-\pi,\pi])$
\cite{Ferrari90:4598,Silberbauer92:7355,Gudmundsson95:16744,PhysRevB.105.155302,PhysRevB.106.115308}.
$\sigma\in\{\uparrow,\downarrow\}$ is the quantum number for the $z$-component of the electron spin,
and all quantum numbers of the Ferrari states are included in $\bm{\alpha}$, which can be viewed as
a subband index.

The Ferrari electron states satisfy the commensurability condition for the competing length scales in
the system, the magnetic length $l$ and the superlattice length $L$, that can be expressed as
$B{\cal A} = BL^2 = pq\Phi_0$
in terms of the unit magnetic flux quantum, $\Phi_0 = hc/e$, and the integers $p$ and $q$
\cite{Hofstadter76:2239,Ferrari90:4598,Silberbauer92:7355,Gudmundsson95:16744}.
Each Landau-band in the energy spectrum will be split into $pq$ subbands. The commensurability
condition can be expressed in different ways \cite{Hofstadter76:2239,Ferrari90:4598,PhysRevB.106.085140},
but it stems from the fact that spatial translations
by superlattice vectors in the external magnetic field gather Peierls phase and have to be replaced
by magnetotranslations.

The total block Hamiltonian (\ref{Hinitial}), for both the electrons and cavity photons, is diagonalized in each
iteration of the DFT scheme in the TP basis (\ref{TP}) and the resulting states for the 2DEG-cavity system are noted by $|\bm{\beta\theta}\sigma)$
together with their wavefunctions (orbitals) $\psi_{\bm{\beta\theta}\sigma}(\bm{r})=\langle\bm{r}|\bm{\beta\theta}\sigma)$.
Important is here to note that in each DFT iteration the electron spin and the current densities are varying
and thus also the para- and the diamagnetic electron-photon interactions (\ref{e-g}) together with the
Coulomb exchange-correlation potentials and functionals.

The expressions for the current and electron densities are given in Appendix \ref{e-TE011}. The mean 
photon number is calculated by defining the photon number operator 
\begin{equation}
      N^{\bm \theta}_\gamma = a^\dagger_\gamma a_\gamma
\end{equation}
in each point, ${\bm \theta}$, in the 1st Brillouin zone. The matrix of the photon number operator 
is assembled in the $|\bm{\alpha\theta}\sigma n\rangle$ basis. The mean photon number is then at each ${\bm \theta}$
\begin{equation}
      \langle a^\dagger_\gamma a_\gamma \rangle^{\bm{\theta}} =
      \mathrm{Tr}\left\{\rho^{\bm \theta}W^{{\bm\theta}\dagger} N^{\bm\theta}_\gamma W^{\bm\theta} \right\}
\end{equation}
with 
\begin{equation}
      \rho^{\bm{\theta}}_{\bm{\alpha}\sigma,{\bm{\beta}}\sigma'} = f\left(E_{\bm{\alpha\theta}\sigma} - \mu \right)
      \delta_{\bm{\alpha},\bm{\beta}}\delta_{\sigma,\sigma'}
\end{equation}
the diagonal density matrix for the interacting 2DEG cavity-photon system in the 
$\{|\bm{\alpha\theta}\sigma)\}$ basis, and $E_{\bm{\alpha\theta}\sigma}$ is the corresponding
energy spectrum. $W$ is the unitary transformation between the $\{|\bm{\alpha\theta}\sigma n\rangle\}$ and the
$\{|\bm{\alpha\theta}\sigma)\}$ bases, and $f$ is the equilibrium Fermi distribution. 
The total photon number $N_\gamma$ is the average of $N^{\bm{\theta}}_\gamma$ over the 1st Brillouin zone. 
The orbital and the spin magnetization
are calculated from the current density and the spin polarization, respectively
\cite{Gudmundsson00:4835,PhysRevB.106.115308}.

%

\section{Results}
\label{Results}
In contrast to a QEDFT approach with an explicit functional describing the electron-photon interactions
\cite{flick2021simple,PhysRevB.106.115308}, where no photon replicas of electron states appear in the energy
spectra, the QED-DFT-TP formalism brings back the cavity photon replicas in a manner comparable to what happens
in models where an exact numerical diagonalization has been used for the respective interactions in a truncated
Fock space \cite{Gudmundsson:2013.305,ANDP:ANDP201500298}. 
Important is though to have in mind the inherent differences between the many-body states of a 
Fock space and the mean-field type of the single-electron DFT states associated with the electron orbitals. 

\begin{figure}[htb]
	\includegraphics[width=0.39\textwidth]{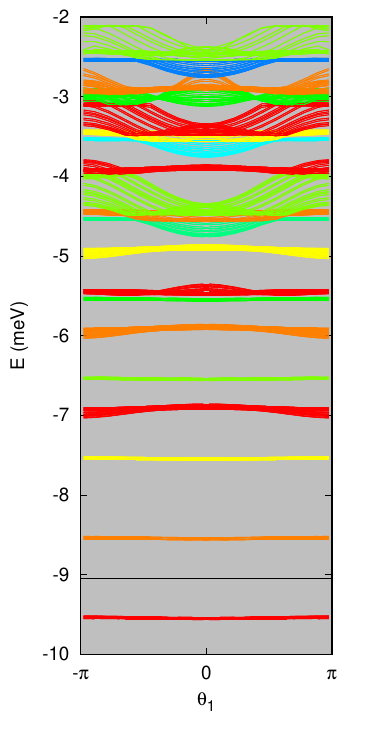}
\caption{The energy bandstructure projected on the $\theta_1$ direction in the 1st Brillouin zone
         for $N_\mathrm{e} = 2$, and $pq=2$. The color of the bands indicates their photon content with red
         for 0 and blue for 12. The chemical potential $\mu$ is shown by the horizontal black line.
         Due to the low magnetic field, $B\approx 0.827$ T, the spin splitting of the bands is not clearly visible on
         the energy scale used. $E_\gamma = \hbar\omega_\gamma = 1.00$ meV, $g_\gamma = 0.001$, $L=100$ nm,
         and $T= 1$ K.}
	\label{EH-pq2-Ne2}
\end{figure}

A QED-DFT-TP spectrum is shown in Fig.\ \ref{EH-pq2-Ne2} for two electrons in each dot, $N_\mathrm{e} = 2$,
and $pq=2$. The two-dimensional spectrum is projected on the $\theta_1$ direction in reciprocal space.
The photon content of the Landau subbands is encoded in their color with red for zero, or a very
low photon number, and deep blue for 12 photons. The low electron-photon coupling $g_\gamma = 0.001$ results in
the total average photon number $N_\gamma \approx 0.00604$ for the two electrons. The low external magnetic field
$B\approx 0.827$ T and the small effective $g^*=-0.44$ lead to a spin singlet with no enhancement of the exchange
energy, and the spin splitting between the two orbital states is thus not discernible on the energy scale of the
figure. The chemical potential indicated with a black horizontal line in the figure is $\mu \approx -9.038$ meV 
and the total energy per dot or cell is $E_\mathrm{tot}\approx -21.96$ meV. 

As $g_{\gamma}$ is small, we can identify photon replicas of the
two states below the chemical potential located at almost regular intervals of $\hbar\omega_\gamma = 1.00$ meV
above them (see the first photon replica orange subbands around -8.5 meV and the second replica
yellow subbands around -7.5 meV).
The higher energy spectrum displays complex structures hinting at resonances and interactions.
Below, they will
be analyzed with more details for situations that bring them closer to the chemical potential of
the relevant system. The photon content in each subband is constant, independent of $\bm{\theta}$, as
the electron-photon interaction Hamiltonian (\ref{e-g}) has no explicit spatial dependence. It is a
functional of the charge and current densities and fits very well into the DFT formalism used.

\begin{figure*}[htb]
	\includegraphics[width=0.48\textwidth,bb=20 50 400 300]{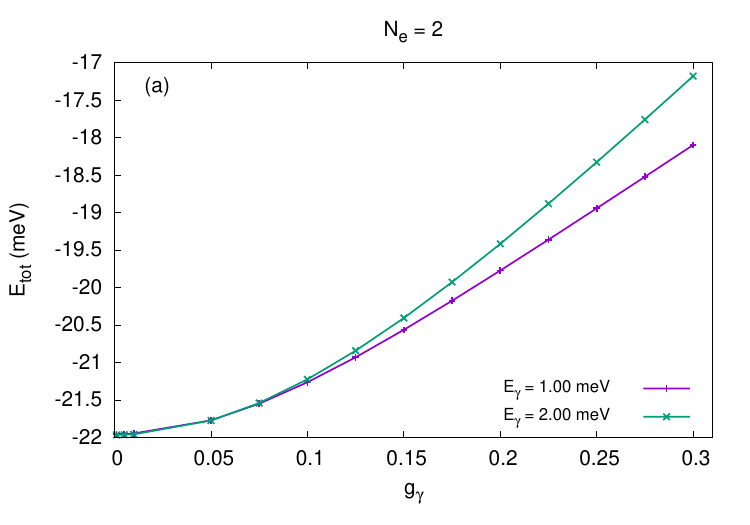}
	\includegraphics[width=0.48\textwidth,bb=20 50 400 300]{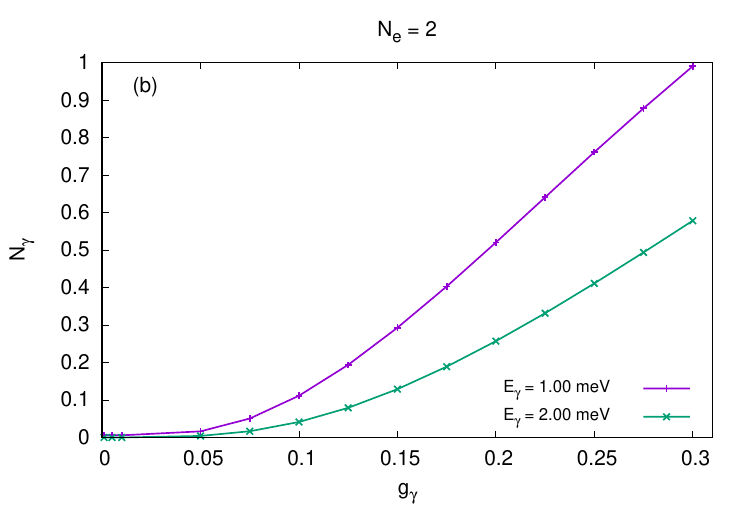}\\
	\vspace*{-0.3cm}
	\includegraphics[width=0.48\textwidth,bb=20 00 400 300]{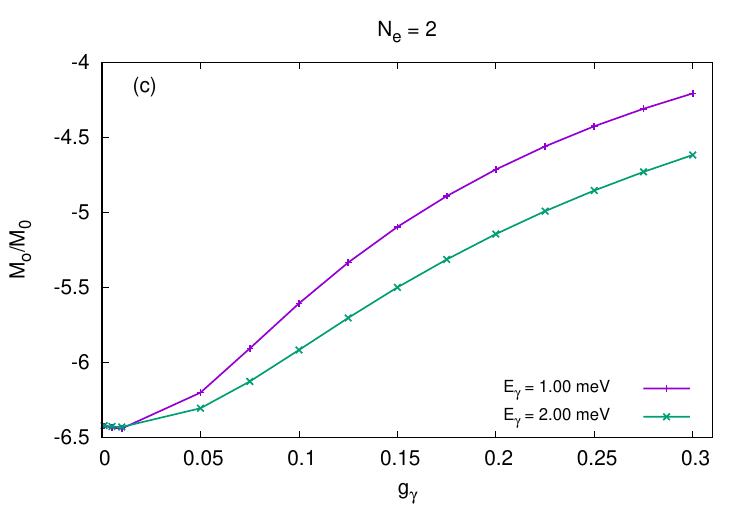}
	\includegraphics[width=0.48\textwidth,bb=20 00 400 300]{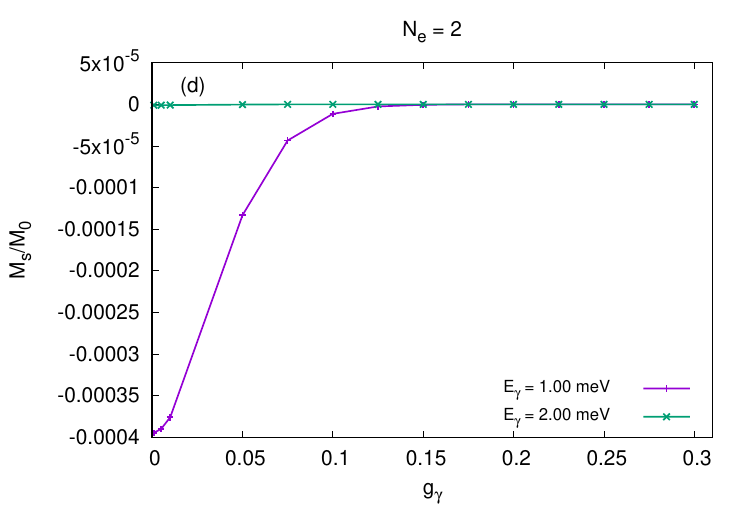}\
	\caption{The total energy (a), the mean number of photons (b), the orbital magnetization (c),
	         and the spin magnetization (d) as functions of the dimensionless electron-photon
	         coupling constant $g_\gamma$ for $N_\mathrm{e} = 2$, and $pq=2$. $L=100$ nm,
	         $T= 1$ K, and $M_0=\mu_\mathrm{B}^*/L^2$.}
	\label{pq2-Ne2-gvar}
\end{figure*}

As will become clear below, the simplicity of the system with 2 electrons in a quantum dot
at low magnetic field and electron-photon coupling makes it ideal to explore what happens
when the electron-photon coupling is increased. In Fig.\ \ref{pq2-Ne2-gvar} the evolution
of the total energy $E_\mathrm{tot}$, the total mean photon number $N_\gamma$, the orbital
$M_o$, and the spin magnetization $M_s$ with increasing $g_\gamma$ for two different values
of the cavity photon energy $\hbar\omega_\gamma$ are presented.

Even for 2 electrons the results are nontrivial as the para- and diamagnetic electron-photon interactions
influence the charge and the current densities that themselves enter the expressions
for the interactions (\ref{e-g}) and moreover, higher order photon- and multiphoton
transitions are included in the QED-DFT-TP formalism. For the selected parameters,
the total energy is higher for the higher photon energy as $g_\gamma$ increases, but at
the same time the number of higher energy photons is lower. The higher energy photons 
polarize the system less effectively, so the orbital magnetization shows a
corresponding effect with respect to the photon energy as the mean photon number, but 
the curvature of $N_\gamma$ and $M_o$ with respect to $g_\gamma$ differs. 
As was realized for the QEDFT formalism \cite{PhysRevB.106.115308}, an increase in
$g_\gamma$ tends to decrease the exchange forces. Here, we see the very small 
Coulomb-exchange contribution to spin magnetization $M_s$ effectively killed by an
increasing electron-photon interaction, i.e.\ the electron-photon interactions force
the 2 electrons into a singlet spin state.

\begin{figure*}[htb]
	\includegraphics[width=0.48\textwidth,bb=20 50 400 300]{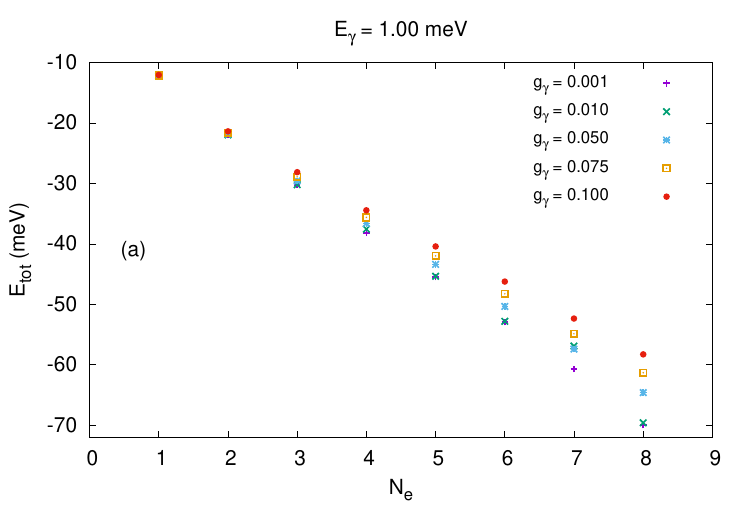}
	\includegraphics[width=0.48\textwidth,bb=20 50 400 300]{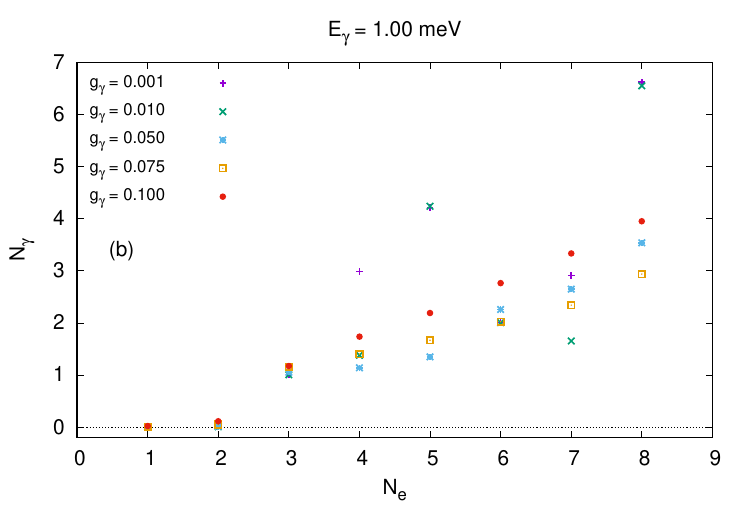}\\
	\vspace*{-0.3cm}
	\includegraphics[width=0.48\textwidth,bb=20 00 400 300]{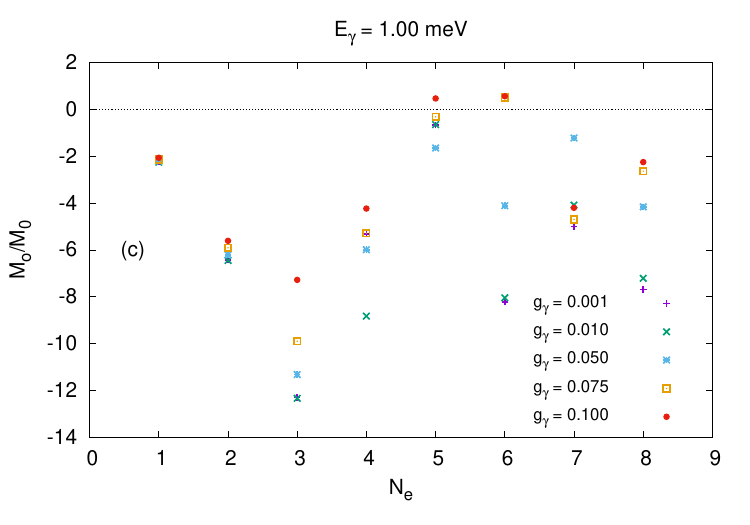}
	\includegraphics[width=0.48\textwidth,bb=20 00 400 300]{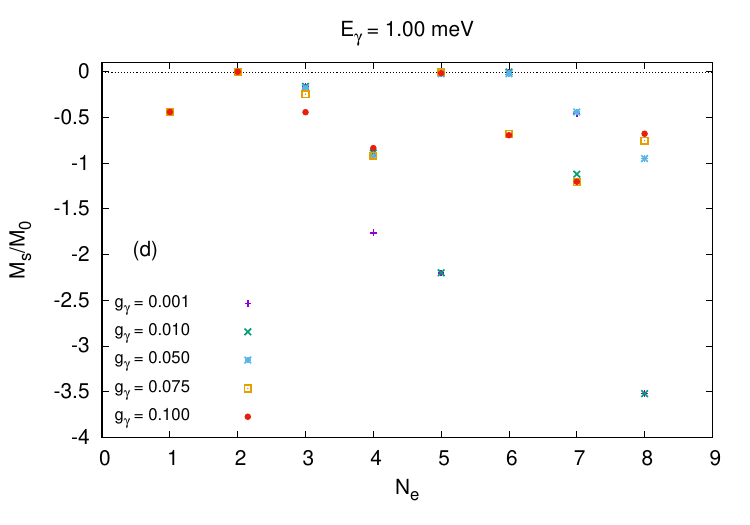}\
	\caption{The total energy (a), the mean number of photons (b), the orbital magnetization (c),
	         and the spin magnetization (d) as functions of the number of electrons $N_\mathrm{e}$
	         for $E_\gamma = 1.00$ meV, and 5 values of the dimensionless electron-photon coupling constant $g_\gamma$.
	         $pq=2$, $L=100$ nm, $T= 1$ K, and $M_0=\mu_\mathrm{B}^*/L^2$.}
	\label{pq2-Ne}
\end{figure*}

As expected, the dependence of the mean values presented in Fig.\ \ref{pq2-Ne2-gvar}
on the electron number $N_\mathrm{e}$ becomes much more complex as then the shell structure
of the quantum dots, or filling factor effects of the modulated 2DEG, come into play.
This is evident in Fig.\ \ref{pq2-Ne} that displays the mean values as functions of
$N_\mathrm{e}$ for several values of $g_\gamma$.

The shell structure, or the filling factor, effects strongly modify both the orbital and the 
spin magnetization as screening and exchange effects play a paramount role in the determination
of the charge and the current densities. Importantly, the electron-photon interactions introduce photon
replica states (see Fig.\ \ref{EH-pq2-Ne2}) into the bandstructure, that complicate further the
shell or the subband structure of the system. For low $g_\gamma$ the coupling of the replica bands
or states is low and many iterations can be needed in the calculations in order to obtain
converged results. Opposite, for high $g_\gamma$ the electron-photon interactions effectively 
subdue Coulomb exchange and correlation effects and generally fewer iterations are needed to reach
convergence. 

Here, we only present how equilibrium quantities and measurables depend on the photon energy
and the electron-photon interactions. No information about time-dependency of transitions is
available in these static calculations, but our experience with the time evolution of small
open electron-photon systems within
the framework of exact numerical diagonalization tells us that some of the converged states 
found in our equilibrium self-consistent calculation would only be reachable in a long time
in time-dependent calculation for an open system \cite{doi:10.1002/andp.201900306}.

\begin{figure*}[htb]
	\includegraphics[width=0.48\textwidth,bb=20 50 400 300]{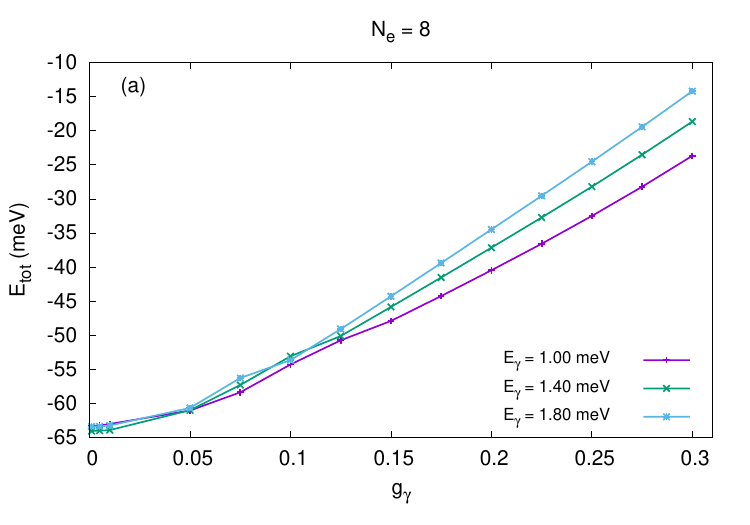}
	\includegraphics[width=0.48\textwidth,bb=20 50 400 300]{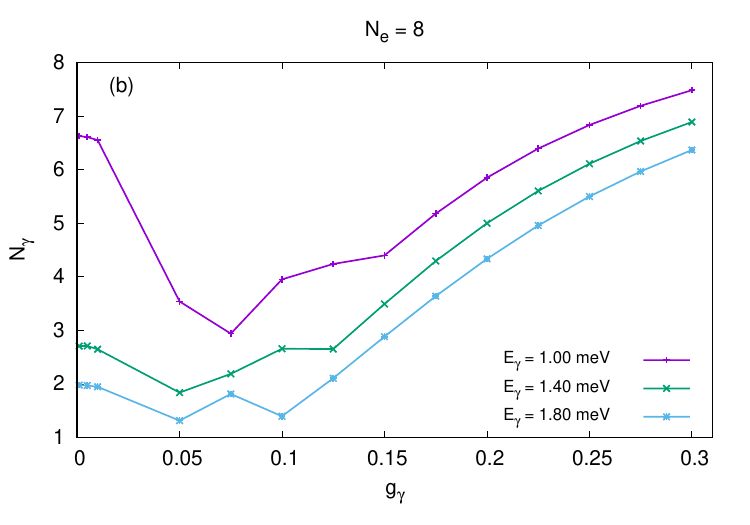}\\
	\vspace*{-0.3cm}
	\includegraphics[width=0.48\textwidth,bb=20 00 400 300]{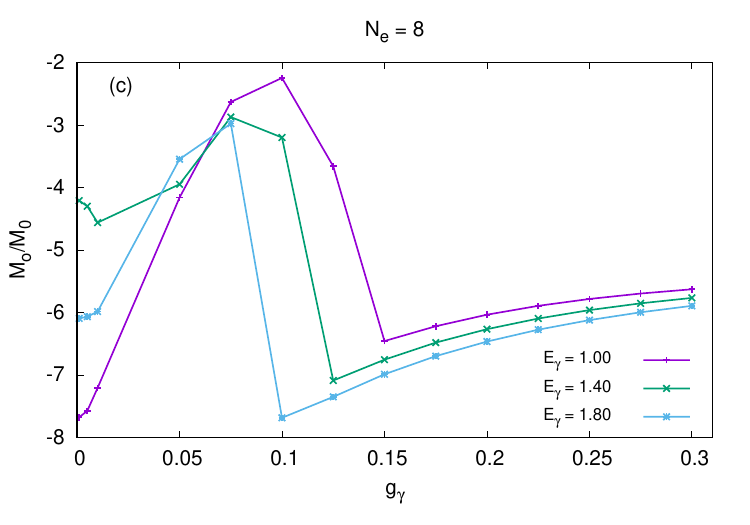}
	\includegraphics[width=0.48\textwidth,bb=20 00 400 300]{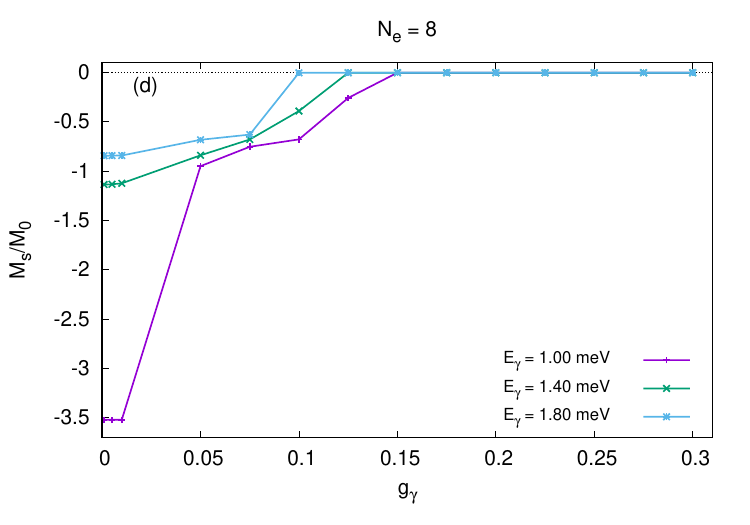}\
	\caption{The total energy (a), the mean number of photons (b), the orbital magnetization (c),
	         and the spin magnetization (d) as functions of the dimensionless electron-photon
	         coupling constant $g_\gamma$ for $N_\mathrm{e} = 8$, and $pq=2$. $L=100$ nm,
	         $T= 1$ K, and $M_0=\mu_\mathrm{B}^*/L^2$.}
	\label{pq2-Ne8-gvar}
\end{figure*}

In Fig.\ \ref{pq2-Ne2-gvar} properties of the system are displayed for two electrons in a
quantum dot or unit cell. Fig.\ \ref{pq2-Ne8-gvar} presents the corresponding averages as
a function of the electron-photon coupling, but now for 8 electrons in a dot for 3 different 
values of the photon energy. In light of what was stated in the previous paragraph it is noticeable
that the results become simpler as $g_\gamma$ surpasses the value 0.15.
In Fig.\ \ref{pq2-Ne8-gvar} it is clearly seen how the electron-photon interactions can effectively
suppress the relatively large Coulomb exchange effects, and how in the intermediate interaction range
the shell or the subband structure determines the averages. 

\begin{figure}[htb]
	\includegraphics[width=0.23\textwidth]{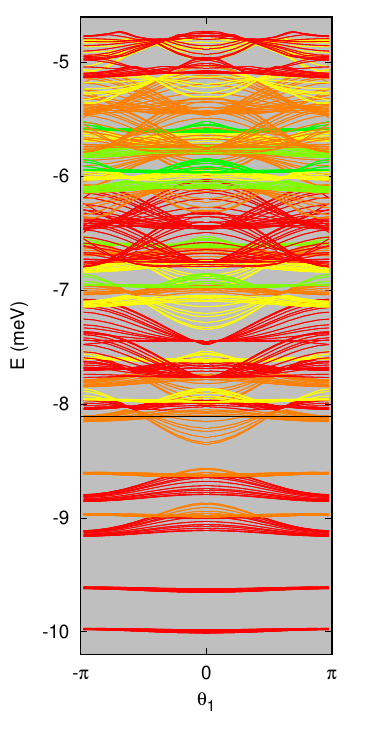}
	\includegraphics[width=0.23\textwidth]{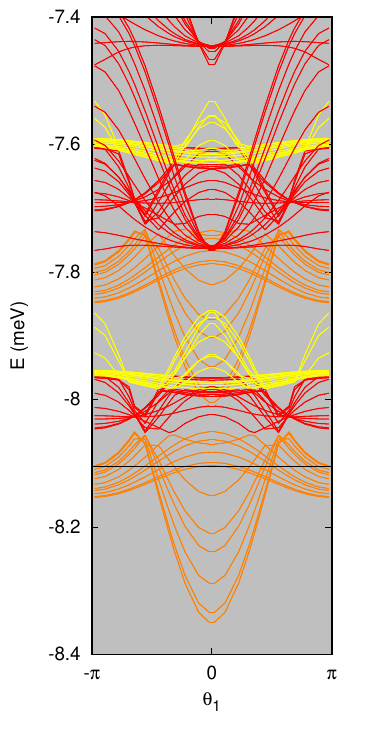}
\caption{The energy bandstructure projected on the $\theta_1 = k_1L$ direction in the 1st Brillouin zone
         for $N_\mathrm{e} = 7$, and $pq=2$ (left), and a Section of the same band structure in
         an energy range around the chemical potential (right).
         The color of the bands indicates their photon content with red
         for 0 and blue for 12. The chemical potential $\mu$ is shown by the horizontal black line.
         An enhanced spin splitting of the bands is seen.
         $E_\gamma = \hbar\omega_\gamma = 1.00$ meV, $g_\gamma = 0.005$, $L=100$ nm, and $T= 1$ K.}
	\label{EH-pq2-Ne7}
\end{figure}

When the electron-photon coupling or the photon energy are varied one can expect resonance conditions
to occur, i.e.\ Rabi resonances like in numerically exact calculations for few electrons 
\cite{PhysRevE.86.046701}. In the present system an analysis of resonances is complicated by the 
two-dimensional shape of the energy subbands in the inverse lattice space and the fact that in 
self-consistent calculations it is not always easy to follow an anticrossing of levels as a single
parameter is varied slightly. But as was seen in Fig.\ \ref{EH-pq2-Ne2}, there are indications of
resonances in the spectra, even for few electrons. The left panel of Fig.\ \ref{EH-pq2-Ne7} shows a large
section of the spectrum for $N_\mathrm{e}=7$, while the right panel shows only a small section close
to the chemical potential (the horizontal black line or plane). The two dimensional energy spectrum is
projected on the $\theta_1$ direction in the inverse space. The color of the bands indicates the 
photon content of the subbands as in Fig.\ \ref{EH-pq2-Ne2}. The right panel shows a clear
anticrossing of bands close to $\mu$, whose structure is delicately dependent on $g_\gamma$,
$E_\gamma$, and $N_\mathrm{e}$. The closeness of this structure to the chemical potential means it
influences most properties of the system. In addition to the Rabi-splitting and anticrossing of
the subbands an enhanced spin splitting is seen.

It is important to realize that the anticrossing displayed in Fig.\ \ref{EH-pq2-Ne7} is over the whole
Brillouin zone, but with varible strength in each point.

\begin{figure*}[htb]
	\includegraphics[width=0.48\textwidth,bb=20 50 400 300]{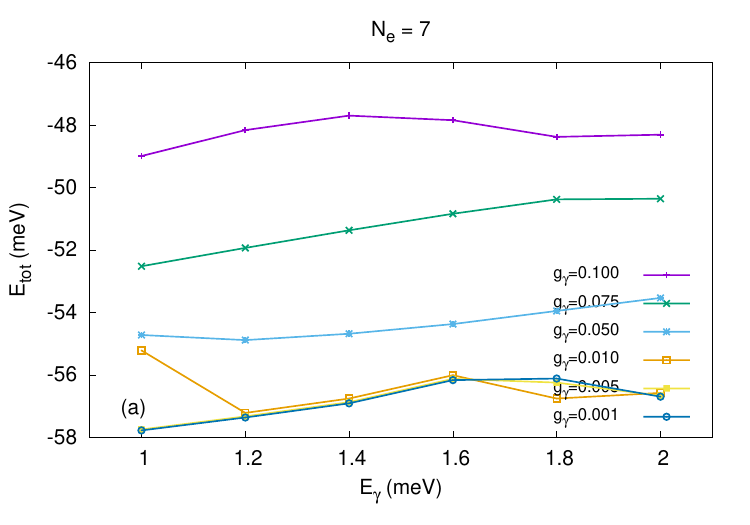}
	\includegraphics[width=0.48\textwidth,bb=20 50 400 300]{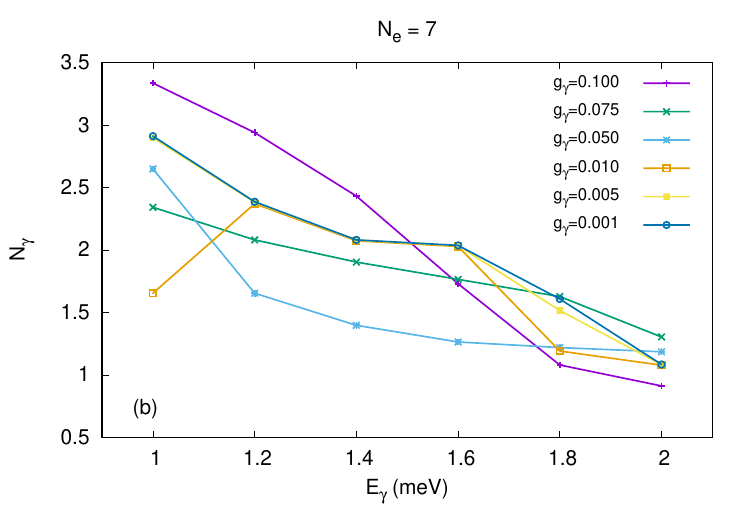}\\
	\vspace*{-0.3cm}
	\includegraphics[width=0.48\textwidth,bb=20 00 400 300]{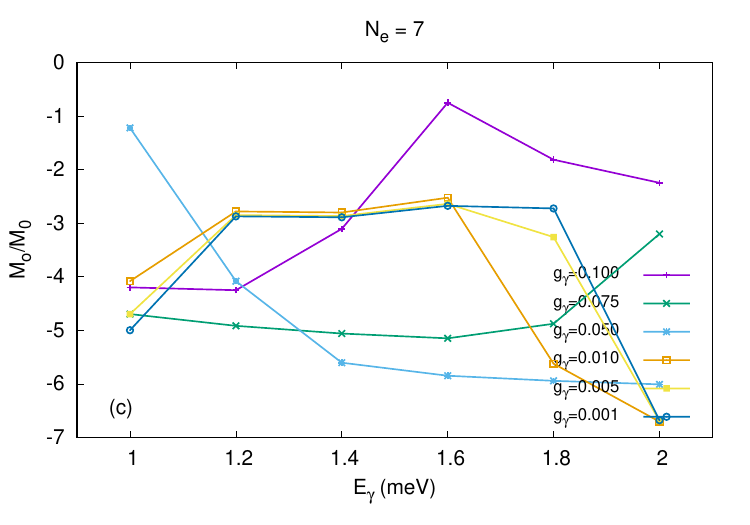}
	\includegraphics[width=0.48\textwidth,bb=20 00 400 300]{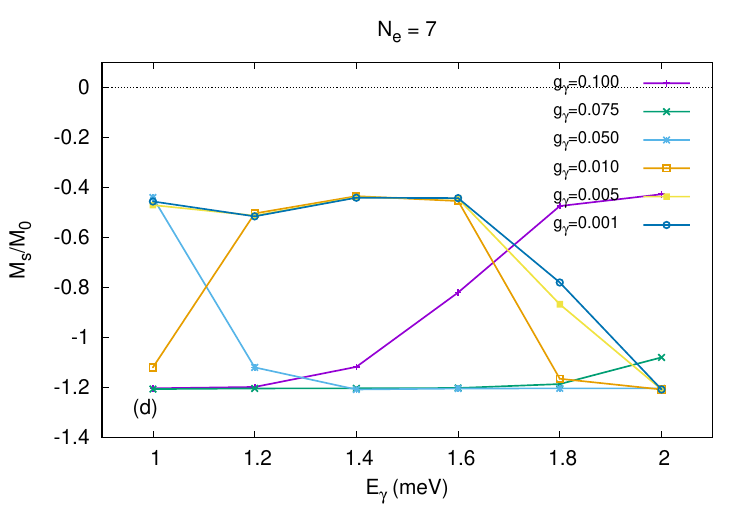}\
	\caption{The total energy (a), the mean number of photons (b), the orbital magnetization (c),
	         and the spin magnetization (d) as functions of the photon energy $E_\gamma$ for
	         $N_\mathrm{e} = 7$, and $pq=2$. $L=100$ nm,
	         $T= 1$ K, and $M_0=\mu_\mathrm{B}^*/L^2$.}
	\label{pq2-Ne7-hwvar}
\end{figure*}

In order to obtain a further insight into the system with 7 electrons in each quantum dot,
we show in Fig.\ \ref{pq2-Ne7-hwvar} the averages as functions of the photon energy
$E_\gamma$ for several values of the electron-photon coupling $g_\gamma$.
The curves for the two or three lowest values of $g_\gamma$ tend to overlap, except for values
of $E_\gamma$ for which the Coulomb exchange interaction creates a difference.
This is connected to the phenomena that the electron-photon interaction tends to reduce
the Coulomb exchange effects and the photon replicas change the shell structure of the
dots.
The spin magnetization in Fig.\ \ref{pq2-Ne7-hwvar} shows transitions between two preferred configurations,
one with approximately 1 odd spin and another one with 3 odd spin $z$-components. Clearly, both the electron-photon
interactions and the photon energy determine the configuration in interplay with the confinement potential and the
effects of the Coulomb interaction.

\begin{figure*}[htb]
	\includegraphics[width=0.48\textwidth]{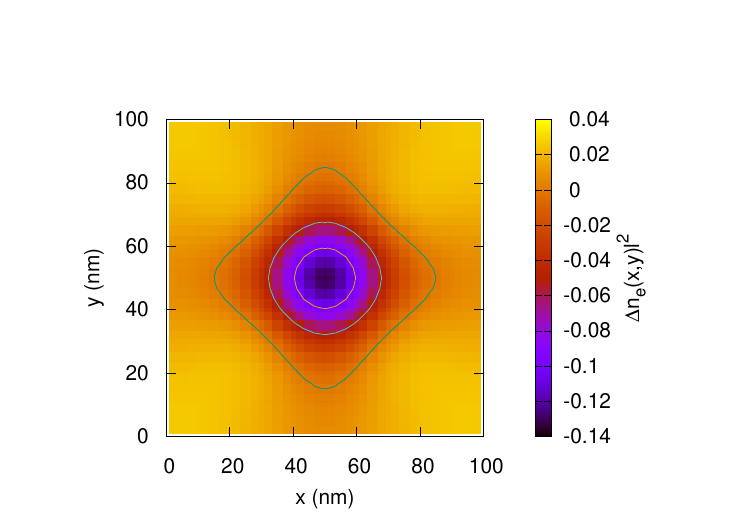}
	\includegraphics[width=0.48\textwidth]{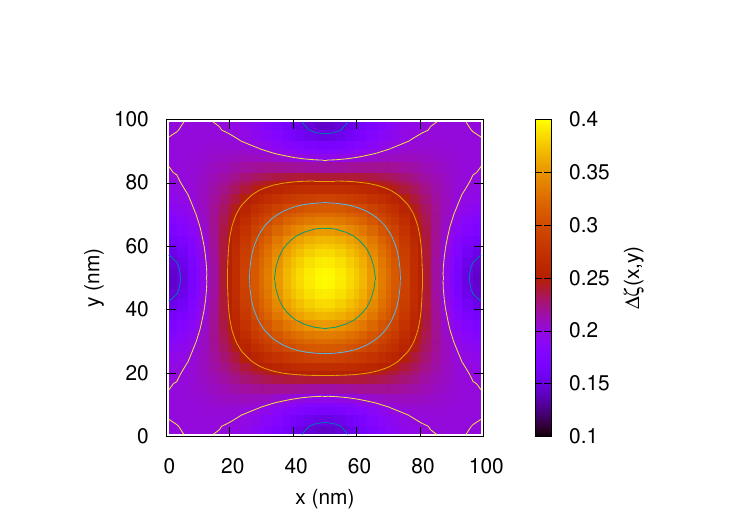}
\caption{The change in the electron density $n_\mathrm{e}(x,y)$ (left), and the electron spin
         polarization $\zeta (x,y)$ (right) when the dimensionless electron-photon coupling constant
         $g_\gamma$ is changed from 0.005 to 0.750. $N_\mathrm{e} = 7$, $pq=2$, and $E_\gamma = 1.0$ meV. 
         $L=100$ nm, and $T= 1$ K}
	\label{pq2-Ne7}
\end{figure*}

We have focused our attention on how the electron-photon coupling affects the global quantities,
like the total energy, the mean photon number, and the orbital and spin magnetization,
but the interaction of the 2DEG  with the cavity photons also leads to local changes or
patterns in the electron properties.
The polarizing effects of the cavity photons are presented in Fig.\ \ref{pq2-Ne7} for both 
the charge and the spin densities of the 2DEG-cavity system. In the left panel it is seen how the polarizing
power of the photons lowers the charge density in the center of each dot and moves it preferably
on the diagonals between the dots to minimize the Coulomb interaction energy.
Note that here the densities are compared for 7 electrons and $E_\gamma = 1.0$ meV, for 
$g_\gamma = 0.005$ and $0.750$. The spin polarization, $\zeta= (n_\uparrow - n_\downarrow)/n_\mathrm{e}$,
in the right panel of Fig.\ \ref{pq2-Ne7} shows a concentration of one spin direction in the
quantum dots as the electron-photon coupling increases, an effect consistent with the information
in Fig.\ \ref{pq2-Ne7-hwvar}(d). Qualitatively corresponding polarization of electron charge was seen in the QEDFT
2DEG model \cite{PhysRevB.106.115308}, but variation in the electron density does imply a variation of
the cavity photon density in the system, an information that was not available in the QEDFT 2DEG model.

\begin{figure*}[htb]
	\includegraphics[width=0.48\textwidth,bb=20 50 380 280]{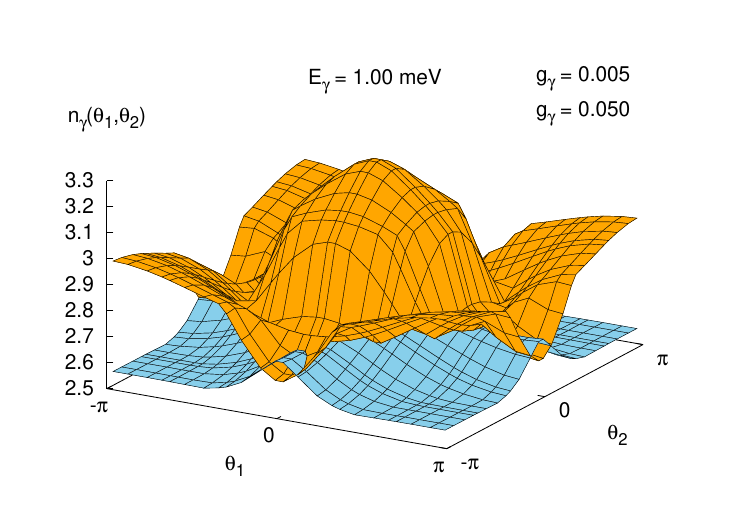}
	\includegraphics[width=0.48\textwidth,bb=20 50 380 280]{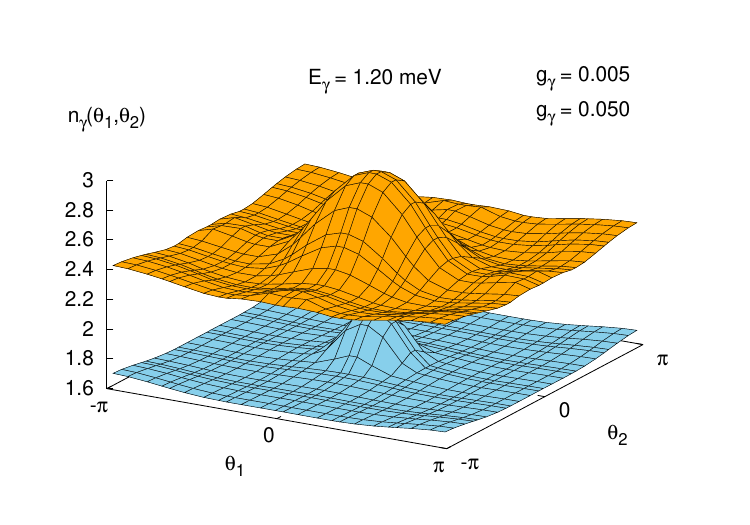}\\
	\includegraphics[width=0.48\textwidth,bb=20 00 380 280]{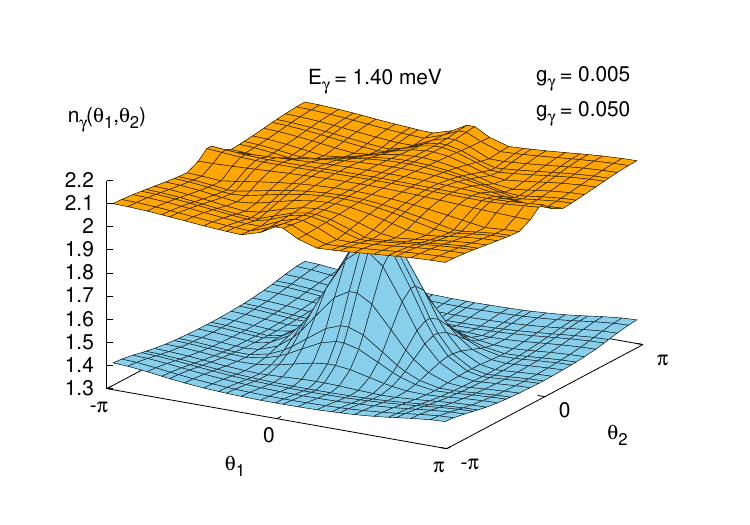}
	\includegraphics[width=0.48\textwidth,bb=20 00 380 280]{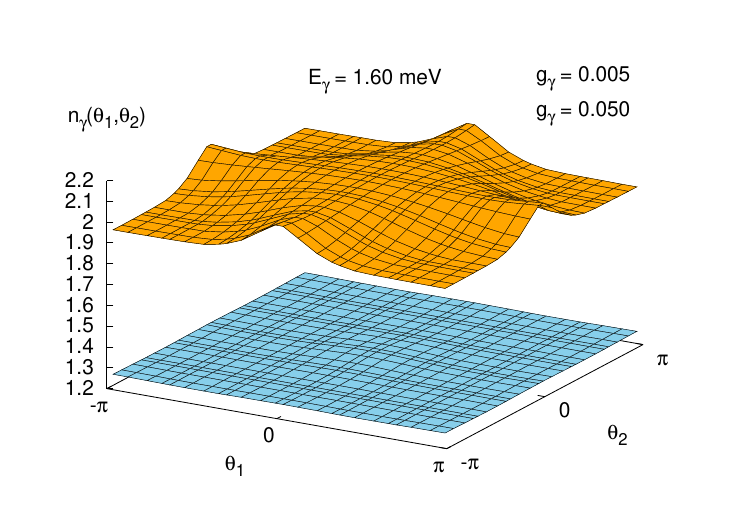}
\caption{The photon density $n_\gamma (\theta_1,\theta_2) = N_{\gamma}^{\bm{\theta}}$ 
         in the 1st Brillouin zone for $N_\mathrm{e} = 7$ and $g_\gamma=0.005$ (orange) and 0.050 (skyblue),
         and four values of the photon energy $E_\gamma$. $pq=2$, $L = 100$ nm, and $T = 1$ K.}
	\label{pq2-Ne7-Nphdens}
\end{figure*}

The photon density, $n_\gamma (\theta_1,\theta_2) = \langle N_{\gamma}^{\bm{\theta}}\rangle$ in the 1st Brillouin zone
of the inverse lattice is displayed in Fig.\ \ref{pq2-Ne7-Nphdens} for two different values of the coupling
constant $g_\gamma$ and four values of the photon energy $E_\gamma$ for the system with $N_\mathrm{e}=7$.

From Fig.\ \ref{pq2-Ne7-hwvar}(b) it is clear that the total photon number does not differ much for 
the two values of $g_\gamma$ in the upper left panel of Fig.\ \ref{pq2-Ne7-Nphdens} for $E_\gamma=1.00$ meV,
but the latter subfigure makes clear that the main difference occurs in the corners of the square unit Brillouin cell.
A further comparison of Figure \ref{pq2-Ne7-Nphdens} and \ref{pq2-Ne7-hwvar}(b) and (d) makes clear
the complex dependence of the photon number, their energy and the coupling constant. Furthermore,
in reciprocal space emerges a nontrivial behavior of the photon density.
The structure of the photon density can be referred back to the energy bandstructure around the
chemical potential. Since the photon content of each subband is constant, large variations in the photon
density for a system with an integer number of electrons in a quantum dot reflect resonances and Rabi 
anticrossing bands around $\mu$. Both the para- (\ref{Hpara}) and the diamagnetic (\ref{Hdia}) parts
of the electron-photon interaction can lead to resonances \cite{2017arXiv170603483G}.

In a superlattice of antidots it is easier to defend a noninteger number of electrons on the average within
each lattice unit.  Mughnetsyan et al.\ have studied the differences in the screening power  and the
magnetic properties of the 2DEG in both the anti and the quantum dot lattice within the QEDFT formalism
for a noninteger number of electrons \cite{Mughnetsyan2023magnetic}.

\section{Conclusions}
\label{Conclusions}
Using a linear basis constructed by a tensor product of one-electron states of the noninteracting Hamiltonian
and the photon states of the number operator we model a 2DEG in a lateral square lattice of quantum dots
placed in a circular cylindrical photon cavity and an external homogeneous magnetic field using a QED-DFT-TP
approach. The total Hamiltonian for the electrons and the photons describable as a photon-block Hamiltonian
with interactions between the blocks determined by the electron-photon interactions,
is diagonalized in each point in the reciprocal space.

After convergence, or self-consistency, is reached in the calculations, the final states are not any more eigenstates
of the photon number operator and the energy subbands of the system have been assigned a constant integer or
fractional photon number. The calculations are performed for an integer number of electrons in each quantum dot
or unit cell of the lattice. The vector potential of the single TE$_{011}$ cavity mode in the long wavelength
approximation has the  same spatial symmetry as the vector potential describing the external magnetic field,
and does not break the symmetry of the original system, but enhances higher order
single-  or multi-photon magnetic processes in both the para- and the diamagnetic electron-photon interactions.

The use of the Coulomb gauge for both the external magnetic field and the cavity field paves the way
to effectively include higher order magnetic processes in the model, and the off-diagonal terms
stemming from the electron-photon interactions together with the photon blocks of the
Hamiltonian of the electron-photon coupling guarantees the inclusion of many-photon processes. We do not use anywhere a
rotating wave approximation for the electron-photon interactions as their antiresonance terms are important
when several processes, virtual or real, close to resonance or not, are active in the system simultaneously.

The photon density in the 1st Brillouin zone of the reciprocal lattice can vary strongly
due to possible anticrossings of subbands with different photon content close to the chemical potential created by
Rabi resonances for certain photon energies. It is more difficult to map the Rabi resonances as can be done
in small confined systems by changing the photon energy or the electron-photon coupling, as in a modulated
2DEG described within a DFT approach the energy subband structure is complicated and depends critically on
both these parameters.

The electron-photon interactions polarize the electron charge as was seen earlier in a QEDFT calculation
for an array of quantum dots and in that process lattice effects are seen depending nontrivially on the
number of electrons in a dot, the photon energy, and their coupling strength to the
electrons \cite{PhysRevB.108.115306}.

Coulomb exchange effects leading to enhanced spin splitting are reduced by the electron-photon coupling.
This phenomena has been observed both for arrays of quantum dots and antidots, though in a slightly
different manner \cite{PhysRevB.108.115306,Mughnetsyan2023magnetic}, but here in the
QED-DFT-TP approach we see a stronger dependence on the photon energy.

In our experience the QEDFT formalism with photon exchange and correlation functionals, but no
explicit photon degrees of freedom \cite{flick2021simple}
gives good qualitative results for both arrays of quantum dots and antidots in an external magnetic field. It can handle multiple cavity photon modes, but it does not give any explicit information about the photon content of the
2DEG  \cite{PhysRevB.108.115306,Mughnetsyan2023magnetic}. In the QED-DFT-TP formalism used in
the present calculations we see a possibility to include in a simpler way both many-photon processes and photon
correlations effects, and the straightforward information about the photon component in the system makes
comparison to calculations using CI approach possible.

\begin{acknowledgments}
This work was financially supported by the Research Fund
of the University of Iceland, and the Icelandic Infrastructure Fund.
The computations were performed on resources
provided by the Icelandic High Performance Computing
Center at the University of Iceland.
V.\ Mughnetsyan and V.\ Gudmundsson acknowledge support
by the Higher Education and Science Committee of Armenia (grant No.~21SCG-1C012).
V.\ Gudmundsson acknowledges support for his visit to the National Taiwan University from the National Science and Technology Council, Taiwan under Grants No.~ NSTC 113-2811-M-002-001 and No.~NSTC 112-2119-M-002-014.
H.-S.\  Goan acknowledges support from the National Science and Technology Council, Taiwan under Grants No.~NSTC 112-2119-M-002-014, No.~NSTC 111-2119-M-002-007, and No.~NSTC 111-2627-M-002-001, and from the National Taiwan University under Grants No.~NTU-CC-112L893404 and No.~NTU-CC-113L891604. H.-S.\ Goan is also grateful for the support from the ``Center for Advanced Computing and Imaging in Biomedicine (NTU-112L900702)\rq\rq  through The Featured Areas Research Center Program within the framework of the Higher Education Sprout Project by the Ministry of Education (MOE), Taiwan, and the support from the Physics Division, National Center for Theoretical Sciences, Taiwan.
J.-D.\ Chai acknowledges support from the National Science and Technology Council, Taiwan 
under Grant No.\ MOST110-2112-M-002-045-MY3.
V.\ Moldoveanu acknowledges financial
support from the Core Program of the National Institute of Materials Physics, granted by the Romanian Ministry
of Research, Innovation and Digitalization under the Project PC2-PN23080202.

\end{acknowledgments}

%
\appendix
\section{The interaction of the 2DEG with a single quantized TE$_{011}$ photon mode of a cylindrical cavity}
\label{e-TE011}
A general TE$_{mnp}$ mode of a circular cylindrical cavity is described by
\begin{equation}
      H_z(r,\phi,z) = B_{mnp} J_m\left(\frac{\chi'_{mn}r}{a}\right)\cos(m\phi)\sin\left(\frac{\pi pz}{d}\right)
\end{equation}
with $a$ and $d$ the radius and the height of the cavity, respectively. $J_m$ is the Bessel function of the
first kind and $\chi_{mn}'$ the $n$th zero of its derivative. The transverse components of the electric field
are then defined by
\begin{equation}
      \bm{E} = \frac{i}{h^2}\omega_{mnp}\mu\left( \bm{e}_z \times \bm{\nabla}_T H_z \right)
\label{Ecav}
\end{equation}
giving for the TE$_{011}$ mode
\begin{equation}
      E_\phi = -B_{011}J_1 \left(\frac{\chi'_{01}r}{a}\right) \sin\left(\frac{\pi z}{d}\right) \frac{i\omega_{011}\mu}{h^2}
\end{equation}
and $E_r = 0$, as $J'_0 = J_1$.  In Eq.\ (\ref{Ecav}) the label $T$ refers to the Cartesian coordinates
perpendicular to $z$, i.e.\ $x$ and $y$, and $B_{011}$ is the strength of the mode, $B_{mnp} $.
The eigenvalue of the Helmholtz equation for the cavity mode is
$h^2 = \omega_{011}^2\mu\kappa - (\pi/d)^2$ which leads to the energy of the TE$_{011}$ mode being
\begin{equation}
      \hbar\omega_{011} = \hbar c\sqrt{\left(\frac{\chi'_{01}}{a}\right)^2 + \left(\frac{\pi}{d}\right)^2}.
\end{equation}
For GaAs parameters for the cavity for the case $a = d = 40\times 10^3$ nm $\hbar\omega_{011}\approx 1.5$ meV,
or 3.1 meV for $a = d = 20\times 10^3$ nm, i.e.\ the radius of the cavity is much larger than the superlattice period $L$.
We thus assume $J_1(x) \rightarrow x/2$ and choose a long wavelength approximation for the vector potential of the cylindrical cavity
\begin{equation}
      \bm{A}_\gamma = \bm{e}_\phi{\cal A}_\gamma\left(a^\dagger_\gamma + a_\gamma \right)\left(\frac{r}{l} \right).
\label{A-TE011}
\end{equation}
This choice is consistent with
\begin{equation}
      \bm{\nabla}\times\bm{A} = \bm{e}_z\frac{1}{r}\left\{\frac{\partial}{\partial r}rA_\phi -
      \frac{\partial A_r}{\partial\phi}\right\} = \bm{e}_z\frac{1}{r} \frac{\partial}{\partial r}rA_\phi,
\end{equation}
and
\begin{equation}
      \bm{E} = -\frac{1}{c}\partial_t \bm{A}.
\end{equation}

In order to evaluate the integrals defining the electron-photon interactions, Eq.\ (\ref{e-g}), we introduce
the notation $\bm{r} = \bm{R} + \bm{x}$ with $\bm{x}$ in the first unit cell of the superlattice and
$\bm{R} = L(m,n) = mL\bm{e}_x + nL\bm{e}_y$ with $n,m\in\mathbf{Z}$ and the lattice length $L$. Furthermore,
$\bm{e}_\phi = -\sin\phi\;\bm{e}_x + \cos\phi\;\bm{e}_y$. Then

\begin{widetext}
\begin{align}
	  H_\mathrm{int}^\mathrm{par} = \frac{1}{c}\sum_{\bm{R}}\int_{\bm{{\cal A}}} d\bm{x}\;
	  \bm{J}(\bm{R}+\bm{x})&\cdot\bm{A}_\gamma (\bm{R}+\bm{x})
	  = \frac{1}{c}\sum_{\bm{R}}\int_{\bm{{\cal A}}} d\bm{x}\; \bm{J}(\bm{R}+\bm{x})
	  \cdot\frac{|\bm{R}+\bm{x}|}{l}\left(\frac{-nL-y}{|\bm{R}+\bm{x}|},\frac{mL+x}{|\bm{R}+\bm{x}|}\right)
	  {\cal A}_\gamma \left(a^\dagger_\gamma + a_\gamma \right)\nonumber\\
	  &= \frac{1}{c}\left\{\frac{(2\pi)^2}{{\cal A}}\delta^G(\bm{G})  \right\}
	  \int_{\bm{{\cal A}}} d\bm{x}\;
	  \left[-J_x(\bm{x})\left(\frac{y}{l}\right) + J_y(\bm{x})\left(\frac{x}{l}\right)\right]
	  {\cal A}_\gamma \left(a^\dagger_\gamma + a_\gamma \right),
\label{Hpara}
\end{align}
and
\begin{align}
      H_\mathrm{int}^\mathrm{dia} &= \frac{e^2}{2m^*c} {\cal A}^2_\gamma \left(a^\dagger_\gamma + a_\gamma \right)^2 \sum_{\bm{R}}\int_{\bm{{\cal A}}} d\bm{x}\;
      n_\mathrm{e}(\bm{R}+\bm{x})A^2_\gamma(\bm{R}+\bm{x})\nonumber\\
      &= \frac{e^2}{2m^*c} {\cal A}^2_\gamma \left(a^\dagger_\gamma + a_\gamma \right)^2 \left[\left\{\frac{(2\pi)^2}{{\cal A}}\delta^G(\bm{G})  \right\}
      \int_{\bm{{\cal A}}} d\bm{x}\; n_\mathrm{e}(\bm{x})\left(\frac{x^2+y^2}{l^2}\right) -
      \left\{N_\mathrm{e}\frac{(2\pi)^2}{{\cal A}} \frac{\partial^2}{\partial (\bm{G}l)^2}\delta^G(\bm{G})\right\} \right],
\label{Hdia}
\end{align}

\end{widetext}
where $N_\mathrm{e}$ is the number of electrons in a unit cell, and we have used
\begin{equation}
      \sum_{\bm{R}} e^{i\bm{R}\cdot\bm{k}} = \frac{(2\pi)^2}{{\cal A}} \delta^G(\bm{k}).
\label{KronDelta}
\end{equation}
together with ${\cal A} = L^2$.
As the spatial integrals in the electron-photon interactions (\ref{e-g}) lead to the interactions to
be expressed as constants multiplied by combinations of the photon creation and annihilation operators,
we interpret (\ref{KronDelta}) as the conversion of a periodic Dirac-delta function to a Kronecker
delta implying that in the matrix elements of the interactions (\ref{Hpara}-\ref{Hdia}) only the
$\bm{G}=0$ terms contribute. Accordingly, we neglect the last term of (\ref{Hdia}). The constants, $I_x$,
$I_y$, and $N$ in (\ref{e-gIxIyN}) are thus
\begin{align}
      l(I_x + I_y) &=  \frac{m^*}{e}\int_{\bm{{\cal A}}} d\bm{x}\;
	  \frac{l}{\hbar}\left[-J_x(\bm{x})\left(\frac{y}{l}\right) + J_y(\bm{x})\left(\frac{x}{l}\right)\right]
	  \nonumber\\
      {\cal N} &= \int_{\bm{{\cal A}}} d\bm{x}\; n_\mathrm{e}(\bm{x})\left(\frac{x^2+y^2}{l^2}\right),
\end{align}
while the electron current and the number densities are
\begin{align}
	\bm{J}_{i}(\bm{r}) = \frac{-e}{m^*(2\pi)^2}\sum_{{\bm \alpha}\sigma}\int_{-\pi}^{\pi} d\bm{\theta}\;
	\Re&\left\{ \psi_{\bm{\alpha\theta}\sigma}^*(\bm{r})\bm{\pi}_i \psi_{\bm{\alpha\theta}\sigma}(\bm{r}) \right\}\nonumber\\
	&f(E_{\bm{\alpha\theta}\sigma}-\mu),
\label{currD}
\end{align}
for $i=x$ or $y$, and
\begin{align}
	  n_\mathrm{e}(\bm{r})
	  = \frac{1}{(2\pi)^2}\sum_{\bm{\alpha}\sigma}
	  \int^{\pi}_{-\pi}d\bm{\theta}\; \left|\psi_{\bm{\alpha\theta}\sigma}(\bm{r}) \right|^2
	  f(E_{\bm{\alpha\theta}\sigma}-\mu),
\label{ne}
\end{align}
respectively. 

The confidence in the derivation of the Hamiltonians (\ref{Hpara}) and (\ref{Hdia}) for the interaction 
of the mode in a cylindrical FIR-cavity with the 2DEG in the long wavelength (\ref{A-TE011}) should be 
enhanced when the reader realizes that the spatial form of ${\bm A}_\gamma$ is the same as for
the vector potential ${\bm A}$ from which the external homogeneous magnetic field is derived.

%
\frenchspacing
\bibliographystyle{apsrev4-2}
%

\end{document}